Hans-Peter Piepho[1], Ingrid Claß-Mahler[2], Beate Zimmermann[2], Wilfried Hermann[3], Jürgen Schwarz[4], Enno Bahrs[2]

# Change of some cropping systems in a long-term trial comparing different systems: rationale and implications for statistical analysis

## Änderung einiger Systeme in einem Dauerversuch zum Vergleich verschiedener Systeme: Begründung und Implikationen für die statistische Analyse

**Affiliations**

[1]University of Hohenheim, Biostatistics (340c), Stuttgart, Germany.
[2]University of Hohenheim, Farm Management (410b), Stuttgart, Germany.
[3]University of Hohenheim, Agricultural Experiment Station (400), Stuttgart, Germany.
[4]Julius Kühn Institute (JKI) – Federal Research Centre for Cultivated Plants, Institute for Strategies and Technology Assessment, Kleinmachnow, Germany.

**Correspondence**

Correspondence: Hans-Peter Piepho, University of Hohenheim, Biostatistics Unit, Institute of Crop Science, Fruwirthstrasse 23, 70599 Stuttgart, Germany, email: piepho@uni-hohenheim.de

**Abstract**

The project "Sustainable agriculture 4.0 without chemical synthetical plant protection" (NOcsPS) tests a number of cropping systems that avoid the use of chemical synthetical pesticides while at the same time using mineral fertilizers. The experiment started in 2020 (sowing in fall 2019). In 2024 (sowing in fall 2023), some of the cropping systems were modified. Analysis of this experiment may be done using linear mixed models. In order to include the data from 2020-2023 in joint analyses with the data collected for the modified systems from 2024 onwards, the mixed modelling approach needs to be reconsidered. In this paper, we develop models for this purpose. A key feature is the use of network meta-analytic concepts that allow a combination of direct and indirect comparisons among systems from the different years. The approach is first illustrated using a toy example. This is followed by detailed analyses of data from the two trials sites Dahnsdorf and Hohenheim.

**Zusammenfassung**

Das Projekt "Nachhaltigere Landwirtschaft 4.0 ohne chemisch-synthetischen Pflanzenschutz" (NOcsPS) testet eine Reihe von Anbausystemen, die ohne die Nutzung von chemisch-synthetischen Pflanzenschutzmitteln auskommen, wobei mineralische Düngung verwendet werden kann. Das Experiment startete im Jahr 2020 (Aussaat im Herbst 2019). Im Jahr 2024 (Aussaat im Herbst 2023) wurden einige der Anbausysteme modifiziert. Die Analyse dieses Experiments kann mit Hilfe linearer gemischter Modelle durchgeführt werden. Um die in den Jahren 2020 bis 2023 erfassten Daten gemeinsam mit den ab 2024 anfallenden Daten unter Einschluss der modifizierten Anbausysteme gemeinsam auswerten zu können, muss der Analyseansatz auf Basis gemischter Modelle adaptiert werden. In der vorliegenden Arbeit entwickeln wir Modelle für diesen Zweck. Ein wesentliches Merkmal des Ansatzes ist die Nutzung von Konzepten aus dem Bereich der Netzwerk Meta-Analyse, welche die Kombination von direkten und indirekten Vergleichen der Systeme in den verschiedenen Jahren erlaubt. Zunächst wird das Verfahren mittels eines kleinen hypothetischen Datensatzes illustriert. Danach folgen detaillierte Analysen für die Daten aus den beiden Versuchsstandorten Dahnsdorf und Hohenheim.

## Introduction

With the objective of developing an agricultural cropping system that also complies with the European Commission's requirements for reducing the use of plant protection products, the project "Sustainable agriculture 4.0 without chemical synthetical plant protection" (NOcsPS) was launched in 2019. NOcsPS cropping systems avoid the use of synthetic chemical pesticides while at the same time using mineral fertilizers. They are intended to enable yield advantages over organic cropping and environmental advantages over conventional cropping systems (ZIMMERMANN et al., 2021). Four years of data from two experiments show that, depending on the location and crop, higher yields can be achieved in NOcsPS cropping systems than in organic cropping systems and that yield losses compared to conventional systems can be reduced through appropriate cultivation measures (CLAß-MAHLER et al., 2023). However, the results also show that not all measures that would potentially be suitable either for generating higher yields or environmental benefits in agricultural systems increase or develop this potential in a NOcsPS cropping system. As a result, some cultivation measures were subsequently discontinued and adapted in order to induce further improvements and optimization within NOcsPS cropping systems. These changes led to the discontinuation of some of the previously tested cropping systems and the introduction of some new ones.

The NOcsPS experiment started in 2020 (sowing in the fall of 2019; subsequently we only refer to the harvest year when mentioning a year) was continued in an adapted form in 2024 (sowing in the fall of 2023). In order to include the data from 2020-2023 in long-term analyses of the experiment, and to ensure comparability of the data collected for the modified systems from 2024 onwards, the approach to joint statistical analysis of all data needs to be reconsidered.

The two long-term trials in Dahnsdorf and Hohenheim have been analysed by a linear mixed model, which is described in detail in CLAß-MAHLER et al. (2023). With the change of some of the cropping systems at both sites in 2024, the treatment design has become unbalanced. This imbalance is addressed by the linear mixed model via the fitted random main effect for years, which provides an adjustment for the fact that the systems have been tested in different sets over time. In this paper, we explain how that model adjustment works. Moreover, we propose a minor modification of the model, i.e., a change from a random to a fixed main effect for years. The impact and rationale of this slight change will be explained using a toy dataset, as well as the real data from the two trials in Dahnsdorf and Hohenheim.

## The changes of cropping systems and their rationale

The results of the first four years of testing (2020-2023) have shown that some of the expectations associated with the measures implemented in the system trials have not been met. The focus in this context was on the yields of the systems and their crops. The yield performance of the systems tested between 2020 and 2023, which formed the basis for decisions on changes to individual systems, is presented in the "Results" section. Essentially, from the fifth year of the trial (2024) onwards, measures relating to the sowing pattern, fertiliser placement and fertilisation strategy were changed, resulting in adjustments in four systems at each of the two trial sites (also see Table 1):

1.) The sowing technology implemented in the NOcsPS II and NOcsPS III systems in the years 2020-2023, which aims to achieve approximate equidistant sowing (aES), was discontinued from 2024. This eliminated the associated narrower row spacing for soybeans and maize and the reduced sowing rate for all crops within these systems. This decision is based on the fact that the aES did not meet the higher yield expectations. The NOcsPS II and NOcsPS III systems were therefore phased out. They have been replaced by the NOcsPS V and NOcsPS VI systems at the Hohenheim site and by the NOcsPS I+ system at the Dahnsdorf site. These new systems are managed using normal sowing and are supplemented by further adapted measures regarding fertilization and biostimulants.

2.) The Cultan fertilization technology tested in the NOcsPS III system in Hohenheim between 2020 and 2023 was also discontinued in 2024. The Cultan technique does not appear to offer the expected benefits in a NOcsPS cropping system as the more intensive mechanical plant protection measures required in NOcsPS cropping systems impair the Cultan depot. The NOcsPS III system has been replaced by the NOcsPS VI system at the Hohenheim site, which is managed by standard fertilizer placement supplemented by further adapted measures regarding fertilization and biostimulants.

3.) In the area of fertilization, a significant change was made in 2024 with the introduction of organic fertilizer in the organic and in some of the NOcsPS cropping systems. The use of organic fertilizer in the ORG systems in Hohenheim and Dahnsdorf from 2024 onwards is intended to improve the nutrient supply of organic cropping systems and increase their yield potential. Organic fertilizer will be applied to wheat, triticale and maize at the Hohenheim site and maize at the Dahnsdorf site. At both sites, the original ORG systems will be continued as ORG+ systems from 2024 with the fertilization adjustments described. Organic fertilizers will be used in some NOcsPS cropping systems in combination with mineral fertilizer to improve various ecosystem services and promote the principle of a circular economy in agriculture and thus takes into account new forms of fertilizer as well as animal husbandry. At the Hohenheim site, this fertilization strategy is implemented in the new NOcsPS V and NOcsPS VI systems, which replace the NOcsPS II and NOcsPS III systems from 2024 onwards. The NOcsPS IV system is continued with the combination of mineral fertilizer and organic fertilizer and is given the suffix "+". No organic fertilizer is used in NOcsPS systems at the Dahnsdorf site.

4.) In the area of nutrient supply, some changes were also made in the organic and the NOcsPS systems from 2024. Overall, the amount of fertilizer will continue to be determined on the basis of the German Fertilizer Ordinance. The only change concerns the calculation of expected yields. For NOcsPS and organic cropping systems, these were originally assumed to be 20% and 50% below the yield values of conventional systems. Since the trial results showed significantly lower yield reductions in some cases, the yield expectations for NOcsPS and organic cropping systems will be calculated from 2024 onwards on the basis of the yield results of the individual systems in the previous trial years. At the Dahnsdorf site, different fertilization levels will be tested in NOcsPS cropping systems starting in 2024 to evaluate the effects of nutrient supply on the yield potential of a NOcsPS cropping system. For this purpose, the terminated NOcsPS II system will be continued as the NOcsPS I+ system in addition to the continued NOcsPS I system, assuming higher yield expectations and a higher fertilization level in line with conventional cropping systems.

5.) In addition, the suitability of biostimulants, micronutrients, and algae extracts for promoting plant growth, disease resistance, and yield in NOcsPS cropping systems are being tested by introducing them into the NOcsPS VI system from 2024 and into the NOcsPS IV+ system from 2025.

**Table 1**: System names of the cropping systems tested in the period 2020-2023 (first column), changed system names of the cropping systems tested 2024 (second column) and characteristics of the cropping systems (columns 3-8) tested at the Hohenheim and Dahnsdorf site in 2024. Cells in bold indicate changes from the original systems.

| **Cropping systems tested in 2020-2023** | **Cropping systems tested in 2024** | | | | | | |
|---|---|---|---|---|---|---|---|
| System name | System name | Crop rotation (years) | Chem.-synth. plant protec-tion | Type of fertilizer | Amount of fertilizer[8)] | Seed pattern | Biostimulants, micro nutrients, algae extracts |
| **Hohenheim** | | | | | | | |
| C I-1 | C I-1 | 3[1)] | yes | mineral | 100%[9)] | normal | no |
| C I-2 | C I-2 | 3[2)] | yes | mineral | 100%[9)] | normal | no |
| C II | C II | 6[3)] | yes | mineral | 100%[9)] | normal | no |
| NOcsPS I | NOcsPS I | 6[3)] | no | mineral | **opt.[10)]** | normal | no |
| NOcsPS II | **NOcsPS V** | 6[3)] | no | **min.+org.** | **opt.[10)]** | **normal** | no |
| NOcsPS III | **NOcsPS VI** | 6[3)] | no | **min.+org.** | **opt.[10)]** | **normal** | **yes** |
| NOcsPS IV | **NOcsPS IV+** | 6[4)] | no | **min.+org.** | **opt.[10)]** | normal | **planned for 2025** |
| ORG | **ORG+** | 6[5)] | no | **organic** | **opt.[10)]** | normal | no |
| **Dahnsdorf** | | | | | | | |
| C II | C II | 6[6)] | yes | mineral | 100%[9)] | normal | no |
| NOcsPS I | NOcsPS I | 6[6)] | no | mineral | **opt.[10)]** | normal | no |
| NOcsPS II | **NOcsPS I+** | 6[6)] | no | mineral | **100%[9)]** | **normal** | no |
| ORG | **ORG+** | 6[7)] | no | **organic** | **opt.[10)]** | normal | no |

1) Winter wheat (WW1) – maize – soybean
2) Winter wheat (WW1) – maize – soybean
3) Winter wheat (WW1) – maize – winter triticale – soybean – winter wheat (WW2) - summer barley
4) Winter wheat (WW1) – maize – winter triticale – soybean – winter wheat (WW2) – ryegrass
5) Winter wheat (WW1) – maize – winter triticale – soybean – winter wheat (WW2) – clover grass
6) Winter wheat (WW1) – maize – winter triticale – pea – winter wheat (WW2) - summer barley
7) Winter wheat (WW1) – maize – winter triticale – pea – winter wheat (WW2) – clover grass
8) according to German Fertilizer Ordinance
9) yield expectations for fertilizer calculation are based on yields of conventional cropping systems
10) yield expectations for fertilizer calculation are lower than for conventional cropping systems and are based on field trial results for individual crops and cropping systems in previous years.

## Implications for statistical analysis

Table 2 illustrates the change of systems for the experiment in Dahnsdorf. The change of systems makes the system × year table incomplete. Up to the year 2023 the design was completely balanced, and any treatment mean over years (systems, crops or system × crop combinations) produced in linear mixed model analyses was essentially equal to an arithmetic mean. With the incompleteness induced by the change of systems in the year 2024, arithmetic means across the full year range (2020 to 2024) can no longer be used for comparing all treatments.

**Table 2**: Cropping systems tested in Dahnsdorf from 2020 to 2024.

| System | Year | | | | |
|---|---|---|---|---|---|
| | 2020 | 2021 | 2022 | 2023 | 2024 |
| 1 ORG | × | × | × | × | |
| 2 C II | × | × | × | × | × |
| 3 NOcsPS I | × | × | × | × | × |
| 4 NOcsPS II | × | × | × | × | |
| 5 ORG+ | | | | | × |
| 6 NOcsPS I+ | | | | | × |

**Table 3**: Toy data (means per system and year) for six cropping systems tested in five years.

| System | Year | | | | |
|---|---|---|---|---|---|
| | 2020 | 2021 | 2022 | 2023 | 2024 |
| 1 | 51 | 49 | 50 | 52 | |
| 2 | 93 | 91 | 88 | 87 | 98 |
| 3 | 57 | 53 | 56 | 52 | 63 |
| 4 | 54 | 51 | 56 | 54 | |
| 5 | | | | | 67 |
| 6 | | | | | 63 |

**Table 4**: Estimated system means for toy data in Table 3 computed with three different models. All models have a fixed system effect.

| System | Model | | |
|---|---|---|---|
| | 1: No year effect | 2: Random year effect | 3: Fixed year effect |
| 1 | 50.5 | 51.952 | 52.175 |
| 2 | 91.4 | 91.4 | 91.4 |
| 3 | 56.2 | 56.2 | 56.2 |
| 4 | 53.75 | 55.2 | 55.425 |
| 5 | 67 | 61.194 | 60.3 |
| 6 | 63 | 57.194 | 56.3 |

**Table 5**: Arithmetic means for the six systems in the toy data in Table 3 for different year ranges.

| System | Year range | | |
|---|---|---|---|
| | 2020-2023 | 2024 | 2020-2024 |
| 1 | 50.5 | | |
| 2 | 89.75 | 98 | 91.4 |
| 3 | 54.5 | 63 | 56.2 |
| 4 | 53.75 | | |
| 5 | | 67 | |
| 6 | | 63 | |
| | | | |
| 2 & 3 | 72.125 | 80.5 | 73.8 |

**Table 6**: Standard errors of a difference (SED) for pairwise comparisons among estimated system means based on toy data in Table 3.

| System comparison | Model | | |
|---|---|---|---|
| | 1: No year effect | 2: Random year effect | 3: Fixed year effect |
| 1 - 2 | 2.3359 | 1.3634 | 1.3484 |
| 1 - 3 | 2.3359 | 1.3634 | 1.3484 |
| 1 - 4 | 2.4622 | 1.3967 | 1.3834 |
| 1 - 5 | 3.8931 | 2.7312 | 2.6789 |
| 1 - 6 | 3.8931 | 2.7312 | 2.6789 |
| 2 - 3 | 2.2023 | 1.2492 | 1.2373 |
| 2 - 4 | 2.3359 | 1.3634 | 1.3484 |
| 2 - 5 | 3.8144 | 2.5168 | 2.4747 |
| 2 - 6 | 3.8144 | 2.5168 | 2.4747 |
| 3 - 4 | 2.3359 | 1.3634 | 1.3484 |
| 3 - 5 | 3.8144 | 2.5168 | 2.4747 |
| 3 - 6 | 3.8144 | 2.5168 | 2.4747 |
| 4 - 5 | 3.8931 | 2.7312 | 2.6789 |
| 4 - 6 | 3.8931 | 2.7312 | 2.6789 |
| 5 - 6 | 4.9244 | 2.7933 | 2.7668 |
| | | | |
| Mean SED | 3.3175 | 2.1257 | 2.0930 |

The linear mixed model suggested for the two NOcsPS experiments (Claß-Mahler et al., 2023) comprises a year main effect. The presence of that effect will ensure an adjustment for differences among years in the computation of treatment means. To illustrate this kind of adjustment, consider the hypothetical toy data for six systems tested in five years presented in Table 3. The missing data pattern equals the pattern for Dahnsdorf shown in Table 2, but the design and amount of data are greatly simplified for illustrative purposes. Also, the simulated yield values bear no resemblance with the real data in the Dahnsdorf trial. To demonstrate the effect and operation of adjustment in the computation of system means, the data were simulated to represent an atypical year 2024, in which two of the systems were changed. Notice that the yields of Systems 2 and 3 in year 2024 are substantially elevated compared the preceding years, indicating an above-average year. These toy data were analysed by linear models having a fixed main effect for systems. Model 1 has no further effects, apart from a residual error term. Model 2 has a random year main effect in addition, and Model 3 has a fixed year main effect in addition. Using factors S and Y to denote systems and years, respectively, the three models can be written in symbolic form as follows (PATTERSON, 1997):

Model 1: S : S.Y

Model 2: S : Y + S.Y

Model 3: S + Y : S.Y

In these three equations, the colon separates fixed effects (before the colon) from random effects (after the colon). S and Y represent the main effects for systems and years, and S.Y denotes a system-by-year effect, which in our simplified toy example comprises both system-by-year interaction and error.

The estimated system means obtained with these three models are presented in Table 4. The results demonstrate that with no adjustment for year effects (Model 1), the treatment means are biased. In particular, the means for the Systems 5 and 6 introduced in the year 2024 are identical to the observed 2024 means in Table 3, and they make Systems 5 and 6 look better than they are in comparison to Systems 1 to 4. In particular, they appear preferable to Systems 1, 3 and 4. This seeming superiority is partly due to the (simulated) exceptionally good year 2024, reflected in the elevated yields for Systems 2 and 3 in that year compared to the performance in years 2020 to 2023. Conversely, the Systems 1 and 4, which were replaced in 2024, have relatively low arithmetic means based on Model 1 because they did not get a chance to display their performance in the favourable year 2024.

Fitting a year main effect (Models 2 and 3) leads to an adjustment of means that accounts for the differences between years. An illustration of the calculations involved for Model 3 is given in the Appendix. In particular, Systems 5 and 6 are adjusted downwards, whereas Systems 1 and 4 are adjusted upwards (Table 4). The adjustment is a bit more pronounced with Model 3, in which the year main effect is fixed. Notice that with both Models 2 and 3 the comparison of Systems 5 and 6 with Systems 2 and 3 based on adjusted means (Table 4) leads to differences that are comparable to the actually observed differences in the year 2024 (Table 3). Overall, Models 2 and 3 provide a similar adjustment. By contrast, the unadjusted means obtained with Model 1 (Table 4) are not commensurate with the differences observed in the year 2024 (Table 3). It is also noteworthy that for Systems 2 and 3, which are the only systems tested in all years, the means are identical under all three models and equal to arithmetic means over the full range or years from 2020 to 2024. These two treatments can be regarded as the common reference for the remaining four systems. It is these two systems remaining in the trial throughout which permit an indirect comparison of Systems 1 and 4 with the new Systems 5 and 6, even though they were never tested together in the same year. The comparison is an indirect one, with the Systems 2 and 3 serving as a common reference, as in network meta-analysis (ADES et al., 2024).

For further illustration, we use the raw data in Table 3 to work out an adjustment of system means from first principles. As will be seen, these adjustments lead to adjusted system means as per Model 3 as presented in Table 4. Table 5 shows arithmetic means for the systems computed for the years 2020 to 2023 and for the years 2020 to 2024. In addition, we reproduce the system means from year 2024. An adjustment may be obtained based on the Systems 2 and 3, which were tested in all five years. The means of these two systems across the full five years are the obvious reference for adjustment, which can be derived by considering the means of Systems 2 and 3 in the three periods displayed in Table 5. The arithmetic mean across Systems 2 and 3 in years 2020 to 2024 is 73.8. By comparison, the arithmetic mean across these two systems in the year 2024 is 80.5, which is 6.7 units above the arithmetic mean for the whole period 2020 to 2024. Hence, to render the means for Systems 5 and 6, which are observed only in year 2024, comparable to Systems 2 and 3, we need to subtract the value of 6.7, yielding adjusted means 67 – 6.7 = 60.3 for System 5 and 63 – 6.7 = 56.3 for System 6. These are exactly the same as the adjusted means based on a linear model with fixed year main effects (Model 3) computed from the full data reported in Table 3. Similarly, the arithmetic mean across Systems 2 and 3 in the years 2020 to 2023 equals 72.125, which is 1.675 units below the mean of these two systems across the full period (73.8). Hence, the means of Systems 1 and 4 over the years 2020 to 2023 need to be adjusted upwards by adding the value 1.675, yielding adjusted means 50.5 + 1.675 = 52.175 and 53.75 + 1.675 = 55.425, respectively. Again, these adjusted means coincide with those obtained using the full data for Systems 1 and 4 with Model 3 (see Table 4).

We can also use this example to illustrate the related concept of indirect comparisons as used in network meta-analysis (Ades et al., 2024). To do so, we use the mean of Systems 2 and 3 as a reference and consider the indirect comparison of Systems 1 and 6, which were never tested together in the same year and so cannot be compared directly. First, we compute the direct difference of the arithmetic mean of System 1 to the arithmetic mean of Systems 2 and 3 in the year range 2020 to 2023, in which they were jointly tested (Table 5). That difference equals $d_1 = 50.5 - 72.125 = -21.625$. The direct difference of System 6 and the mean of Systems 2 and 3 in the year 2024 equals $d_6 = 63 - 80.5 = -17.5$. The indirect difference between Systems 1 and 6 equals the difference of these two direct differences, i.e., $d_1 - d_6 = -21.625 - (-17.5) = -4.125$. This difference coincides with the difference of adjusted means for Systems 1 and 6 based on Model 3 fitted to the full data (Table 4): 52.175 – 56.3 = –4.125. The same kind of indirect comparison can be made, e.g., for Systems 4 and 5 (calculations not shown), yielding a difference of –4,875, which agrees with the difference of the adjusted means obtained with Model 3 (Table 4).

In analyses performed for the Dahnsdorf and Hohenheim trials up to the year 2023, the year main effect was modelled as random (Claß-Mahler et al., 2023). An alternative is to use a fixed year main effect (see next section). As illustrated in the toy example, this is expected to afford a slightly larger adjustment for year effects than a model with random year main effects. It may be added that when the year main effect is modelled as fixed (Model 3), all inference on treatment comparisons is based on comparisons within years. By contrast, a model with random year main effects (e.g., Model 2) also includes comparisons across years and allows for what is known as 'recovery of inter-year information' (Piepho et al., 2024a), and this explains the somewhat smaller adjustments based on that model observed in Table 4. Our proposed slight change of model (year main effect fixed instead of random) is in agreement with standard practice in network meta-analysis (Piepho, et al., 2024b).

A further consideration is the small number of years. With only five years, the variance for the year main effect is estimated on only four degrees of freedom, rendering the variance estimate rather inaccurate, adversely affecting the weighted combination of intra-year and inter-year information, in which the inverse variances play the role of weights (Yates, 1940). A simple method to decide whether or not to recover the inter-year information is to fit both models, i.e. with fixed and with random year main effect, adjusting the denominator degrees of freedom using the Kenward & Roger (1997) method, and then inspecting the standard errors of a difference (SED) among system means. The SED are computed automatically by the linear mixed model package, using residual maximum likelihood estimate of the variance components. Explicit equations involve matrix expressions (McLean et al., 1991), which are not reproduced here. The SED among estimated system means in Table 6 show that for the toy data the recovery of inter-year information is not warranted because standard errors are smaller for the model with fixed year effects (Model 3).

For the simulated data in Table 3, we computed the mean of all SEDs for comparisons among adjusted system × crop means in the analysis based on all data of the years 2020 to 2024. This yielded a mean SED = 0.888 for fixed year main effects and a mean SED = 0.900 for random year main effects, indicating that the fixed effect assumption is preferable. For the Hohenheim trial, we find mean SED = 0.919 for fixed year main effects and mean SED = 0.918 for random year main effects, which is a very slight edge in favour of random year main effects. For the 2024 analysis, we will use the fixed effects assumption for both trials for consistency. In each subsequent year we will check the SEDs again to decide on the status of the year main effect as fixed or random. As more years of data accumulate, the balance is expected to tip in favour of random year main effects (Model 2) because the associated degrees of freedom for the variance component increase (Möhring et al., 2015). More generally, the adjustment for year effects afforded by both Models 2 and 3 is expected to stabilize with an increasing number of years.

Mean comparisons are facilitated by adding letters as superscripts on the estimated means. Due to the imbalance of the treatment design after the change of some systems in 2024, the SEDs are quite heterogeneous between pairs of systems (Table 6). In Claß-Mahler et al. (2023), we used the LINES option of the GLIMMIX procedure of SAS for mean comparisons. The algorithm used with this option occasionally fails to truthfully report all significant comparisons in the letter display, especially when the SED are very heterogeneous, as is expected when the systems-by-year classification is very incomplete after a change of some systems into new ones. For example, with the toy data and Model 3, it cannot display the significant comparisons between Systems 1 and 3 and between Systems 1 and 4. The algorithm fails because it tries to connect systems that are not significantly different by lines, but with heterogeneous SED occasionally such a line would suppress a significant difference in the display. For a detailed description of this problem, and possible remedies, see Piepho (2004). An alternative in SAS is the %MULT macro (Piepho, 2012), which implements the insert-and-sweep method proposed in Piepho (2004). The same option is also available in R via the 'multcompView' and 'emmeans' packages. Table 7 shows the letter display for the mean comparison based on the toy data using Model 3 in two different forms. One is the simultaneous comparison among all six systems. The other option is to only consider the four current Systems 2, 3, 5 and 6. To obtain this letter display, a grouping factor G is defined with level G='current' for the four current systems and level G='ended' for Systems 1 and 4. The treatment effect is then modelled using the nested structure G/S, where S represents the system, and mean comparisons are stratified by G. Which of these two forms of mean comparison is preferable depends on the objective of the analysis. The change of systems was implemented because the modification is expected to lead to improved performance. It may therefore be of primary interest to compare the current systems. However, in addition one may want to establish whether or not the change of system indeed led to an improvement as expected, which case all six systems need to be compared. Importantly, even if only the current systems are to be compared, analysis should always be based on the complete data, including the data for the terminated systems, because these also provide valuable information about the variance components of random effects, and on the year-related effects. Analysing just subsets of the data is therefore not recommended.

**Table 7**: Letter display for mean comparison among systems for the toy data based on Model 3 for (i) all systems and for (ii) the current Systems 2, 3, 5 and 6 only.

| Group (G) | System (S) | Systems compared§ | |
|---|---|---|---|
| | | All systems | Systems 2, 3, 5 and 6 |
| ended | 1 | 52.2[c] | |
| current | 2 | 91.4[a] | 91.4[a] |
| current | 3 | 56.2[b] | 56.2[b] |
| ended | 4 | 55.4[b] | |
| current$ | 5 | 60.3[b] | 60.3[b] |
| current$ | 6 | 56.3[bc] | 56.3[b] |
| | | | |
| | Mean SED | 2.09 | 2.32 |

§ Means in a column followed by a common letter are not significantly different according to a t-test at the 5% level of significance.
$ New system started in 2024.

**Design and linear mixed model for the Dahnsdorf and Hohenheim trials 2020 to 2024**

The trial layout in both Dahnsdorf and Hohenheim was a strip-plot design with systems randomly allocated to rows within replicates and phases of the system (Patterson, 1964) randomly allocated to columns. The model used for joint analysis across years for data up to 2023 is described in detail in Claß-Mahler et al. (2023). Briefly, the year-specific block effects are represented by the random-effects structure

REP.Y + REP.POW.Y + REP.COL.Y + REP.ROW.COL.Y

where factors REP, ROW, COL and Y represent replicates, rows nested within replicates, columns nested within replicates, and years, respectively. For each random effect, a first-order auto-regressive model (AR1) was fitted to account for serial correlation over years.

Up to the year 2023, before the change of systems, the year-specific treatment structure was modelled as

Model 4: S.P.C.V : Y + S.P.C.V.Y

where factors S, P, C and V represent systems, phase (position within a cycle), crops and varieties. As with our toy example, the key term regarding the adjustment for year effects is the main effect for year (Y). This model corresponds to Model 2 for the toy data. Even though the model had this random main effect in analyses of data up to 2023, it did not lead to an adjustment of the means, because the system-by-year classification was complete. With the change of some systems from 2024 onwards, however, the classification is incomplete. Hence, the adjustment is needed, and this adjustment is automatically effected by fitting this model, as was demonstrated with the analogous Model 2 for the toy data (Table 4).

Analysis by Model 4 recovers the inter-year (between-year) information, because the year main effect is random. If we want to base system comparisons only on intra-year (within-year) comparisons, we need to model the year main effect as fixed. This would correspond to the model

S.P.C.V + Y : S.P.C.V.Y

We initially considered this model but found that its use led to unexpected mean adjustments for some crops. Notice that in our toy example, we only considered the factor S (system). Now, for the real data from the two field experiments, we also need to take the crop (C) into account. It is to be expected, that year effects are crop specific. To make the adjustment crop-specific, and to ensure that no inter-year information is recovered, we replace the fixed year main effect Y with a fixed crossed effect C.Y:

Model 5: S.P.C.V + C.Y : S.P.C.V.Y

This model corresponds to Model 3 for the toy data. Model 5 is our suggested model for joint analyses over years of data arising from the experiments at Dahnsdorf and Hohenheim starting from 2024 onward.

Inspection of the studentized residuals based on Models 4 and 5 indicated that a data transformation was needed to stabilize the variances. Here, we use the square-root transformation throughout. All inference, i.e. fitting of the linear mixed model, calculation of adjusted means and standard errors (SED), confidence intervals and significance tests is done on the transformed scale. For reporting in figures and tables, the adjusted means are transformed back to the original scale. These back-transformed adjusted means can be interpreted as estimates of medians as explained in detail in PIEPHO (2009). In the next section, we therefore refer to these back-transformed means briefly as medians.

## Results for the analysis of the Dahnsdorf and Hohenheim trials 2020 to 2024

The approach developed in this study is applied below to evaluate the yield results collected in the NOcsPS research project for different cropping systems and time periods jointly over a five-year period (2020-2024). The focus is on the yield results from the two trial sites Hohenheim and Dahnsdorf for winter wheat (WW1 and WW2), maize, and legumes (soybeans in Hohenheim and peas in Dahnsdorf). To assess the introduced model adjustment, the model-based medians of the yield results are shown in the following figures and the arithmetic mean values of the yields as computed from the raw data are shown in the tables below. This approach was chosen for both locations.

Without taking system adjustments into account, 2024 there was a general decline in yields compared to previous years at both sites and in all cropping systems, with the exception of maize in Hohenheim and legumes at both sites. This effect, which occurred independently of the system changes and adjustments, is likely due to the warm and humid weather conditions in early summer 2024, since, in contrast to previous years, fungal diseases such as DTR, septoria, yellow and brown rust, mildew and fusarium occurred in both wheat varieties, depending on the stage of development of the plants. For instance, annual precipitation for Hohenheim in 2024 totaled 868.9 mm, 177.8 mm above the long-term average, with the greatest differences occurring in the months of May and June.

In detail, the average yields for winter wheat (WW1) in Hohenheim in the years 2020-2023 (Figure 1 and Table 8) do not differ significantly from each other, with the exception of the ORG system. However, there are differences in the average yields of all systems in the years 2020-2024. It can be assumed that the lower yields in 2024 will further increase this effect. In 2024, the yields of all NOcsPS cropping systems are significantly different compared to the conventional systems. It should also be noted that the conventional cropping system with a six-year crop rotation has a significantly lower yield in WW1 than the conventional cropping system with a three-year crop rotation, possibly due to the less favorable previous crop of WW1. The significantly lower yields of the NOcsPS cropping systems compared to previous years could also be an indication of the changed fertilization strategy with the use of organic fertilizer. However, the unchanged NOcsPS I system also shows a lower yield in 2024. On a five-year average, the current NOcsPS systems NOcsPS V+, V, and VI are inferior to the discontinued systems NOcsPS II, NOcsPS III, and NOcsPS IV+ in terms of yield. Whether this is due to unsuitable modification of the systems or to the new systems not yet being implemented optimally needs to be investigated.

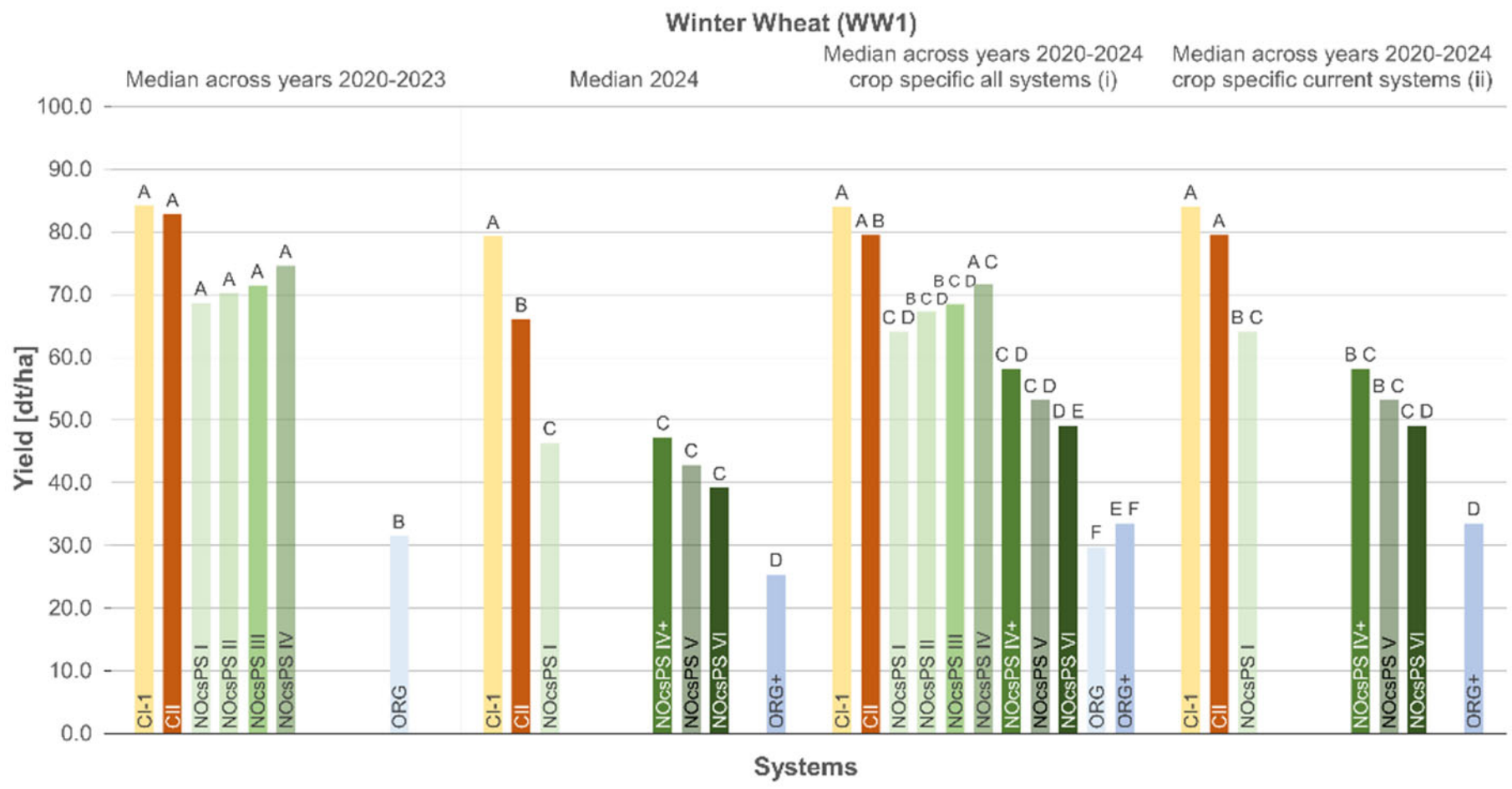

**Figure 1**: Medians of winter wheat yields [dt/ha] across different timeframes, years 2020-2023, year 2024 and years 2020-2024 of the tested systems for Asory in Hohenheim. Letter display for mean comparison among systems for timeframe 2020-2023 data based on Model 4 and for timeframe 2020-2024 data based on Model 5 for (i) all systems and for (ii) the current Systems only.

**Table 8**: Arithmetic means of winter wheat yields [dt/ha] (Asory, Hohenheim) of the single years 2020 to 2024, across years 2020-2023 and 2020-2024 for all tested systems.

| Year | **CI-1** | **CII** | **NOcsPS I** | NOcsPS II | NOcsPS III | NOcsPS IV | **NOcsPS IV+** | **NOcsPS V** | **NOcsPS VI** | ORG | **ORG+** |
|---|---|---|---|---|---|---|---|---|---|---|---|
| 2020 | 80.6 | 87.1 | 74.3 | 70.5 | 73.3 | 76.1 | | | | 23.5 | |
| 2021 | 91.9 | 86.2 | 79.8 | 75.4 | 78.6 | 67.1 | | | | 29.2 | |
| 2022 | 83.5 | 78.8 | 58.5 | 70.0 | 68.3 | 87.5 | | | | 40.1 | |
| 2023 | 60.9 | 57.4 | 65.2 | 66.2 | 67.0 | 69.6 | | | | 34.8 | |
| Across years 2020-2023 | 85.5 | 84.1 | 70.2 | **70.5** | **71.8** | **74.9** | | | | 32.5 | |
| 2024 | 80.3 | 67.0 | 47.0 | | | | 47.9 | 43.6 | 39.7 | | 26.1 |
| Across years 2020-2024 | **84.1** | **79.8** | **65.3** | | | | | | | | |

For winter wheat (Asory), the changes in the significances of all systems to those of the current systems in the years 2020-2024 are apparent and illustrate the influence of the research questions. An evaluation of the changes within the systems is not yet possible after one year; the effects will become clearer in the following years. Nevertheless, it could be shown that the redesigned new systems (NOcsPS IV+, NOcsPS V and NOcsPS VI), based on the evaluation with the model established and applied specifically for this project, reflect the trends in the yields of the individual systems from previous years as well as the trends in the yields of the systems in 2024 in the summarized figure for the years 2020 to 2024. These statements can also be applied to the results for the Dahnsdorf site (Figure 2 and Table 9).

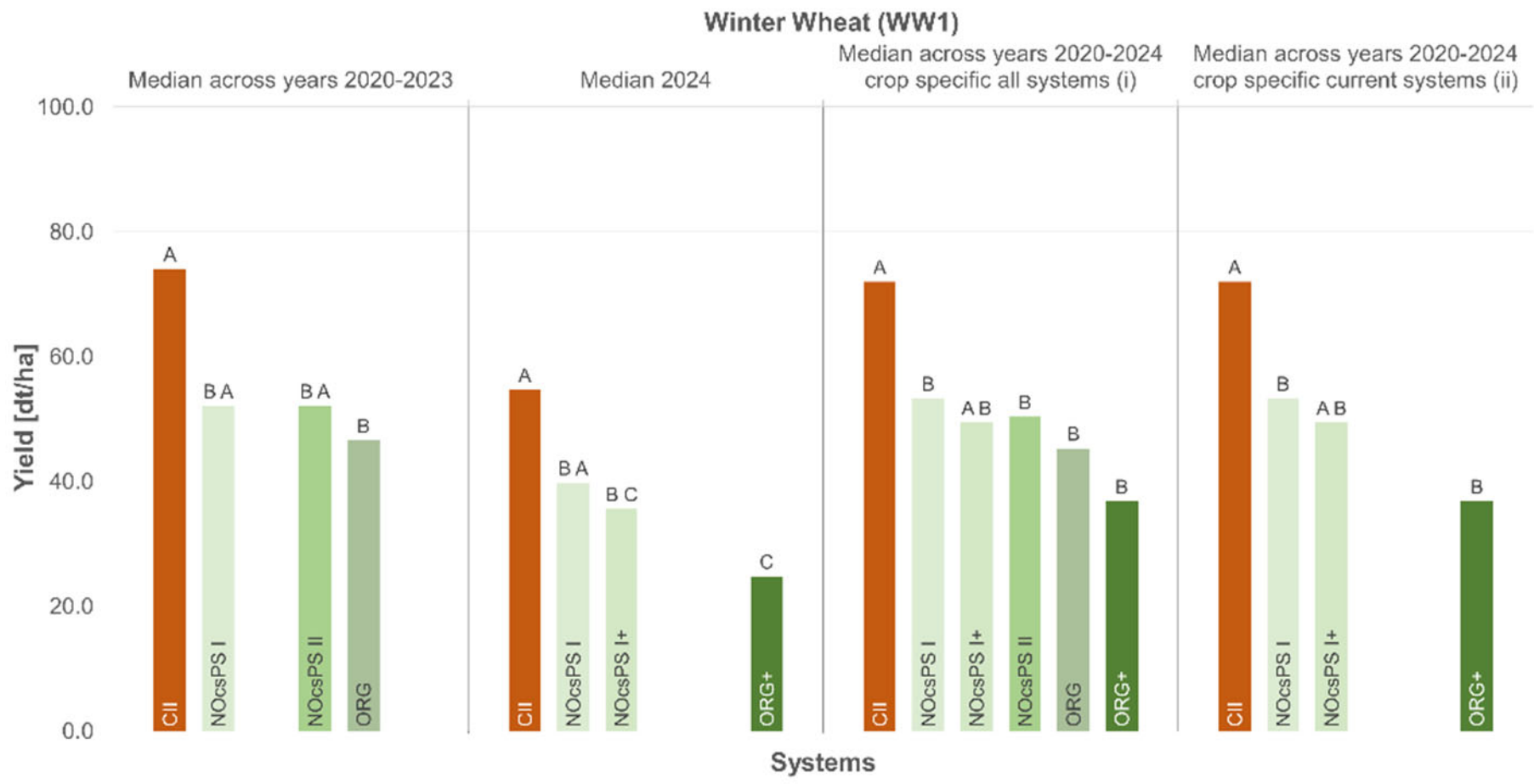

**Figure 2**: Medians of winter wheat yields [dt/ha] across different timeframes, years 2020-2023, year 2024 and years 2020-2024 of the tested systems for Achim in Dahnsdorf. Letter display for mean comparison among systems for timeframe 2020-2023 data based on Model 4 and for timeframe 2020-2024 data based on Model 5 for (i) all systems and for (ii) the current Systems CII, NOcsPS I, NOcsPS I+ and ORG+ only.

**Table 9**: Arithmetic means of winter wheat yields [dt/ha] (Achim, Dahnsdorf) of the single years 2020, 2021, 2022, 2023 and 2024 and across years 2020-2023 and 2020-2024 for all tested systems.

| Year | **CII** | **NOcsPS I** | **NOcsPS I+** | NOcsPS II | ORG | **ORG+** |
|---|---|---|---|---|---|---|
| 2020 | 96.4 | 85.3 | | 75.1 | 67.3 | |
| 2021 | 82.1 | 63.6 | | 59.1 | 51.6 | |
| 2022 | 63.3 | 41.5 | | 45.3 | 42.8 | |
| 2023 | 69.9 | 45.5 | | 42.1 | 41.6 | |
| Across years 2020-2023 | 78.0 | 59.0 | | **55.4** | 50.8 | |
| mean 2024 | 55.0 | 39.8 | 35.9 | | | 24.9 |
| Across years 2020-2024 | **73.4** | **55.1** | | | | |

As for WW1 (Asory, Hohenheim), the average wheat yields from the years 2020-2023 (Figure 3 and Table 10) for RGT Reform are not significantly different from each other, except the ORG system. There are also no differences in the average yields from all systems over the years 2020-2024, except the ORG and ORG+ systems. Nevertheless, it should be noted that the decline in yields in all systems in 2024 led to a yield decrease in the NOcsPS cropping systems with significant differences to the conventional systems. In addition to the seasonal disease pressure mentioned above, the use of organic fertilizer in the NOcsPS IV+, NOcsPS V, and NOcsPS VI systems may have contributed to a yield decline due to a potentially unfavorable nutrient supply to the plants in terms of timing and location. Due to the similarly reduced yield in the unchanged NOcsPS I system, this conclusion must be further observed in subsequent years. However, this result could not be found in Dahnsdorf for RGT Reform, where yields were more stable in 2024 (Figure 4 and Table 11). Apart from the influence of the pathogen load, increased fertilization in the NOcsPS I+ system may have had a positive effect on yield.

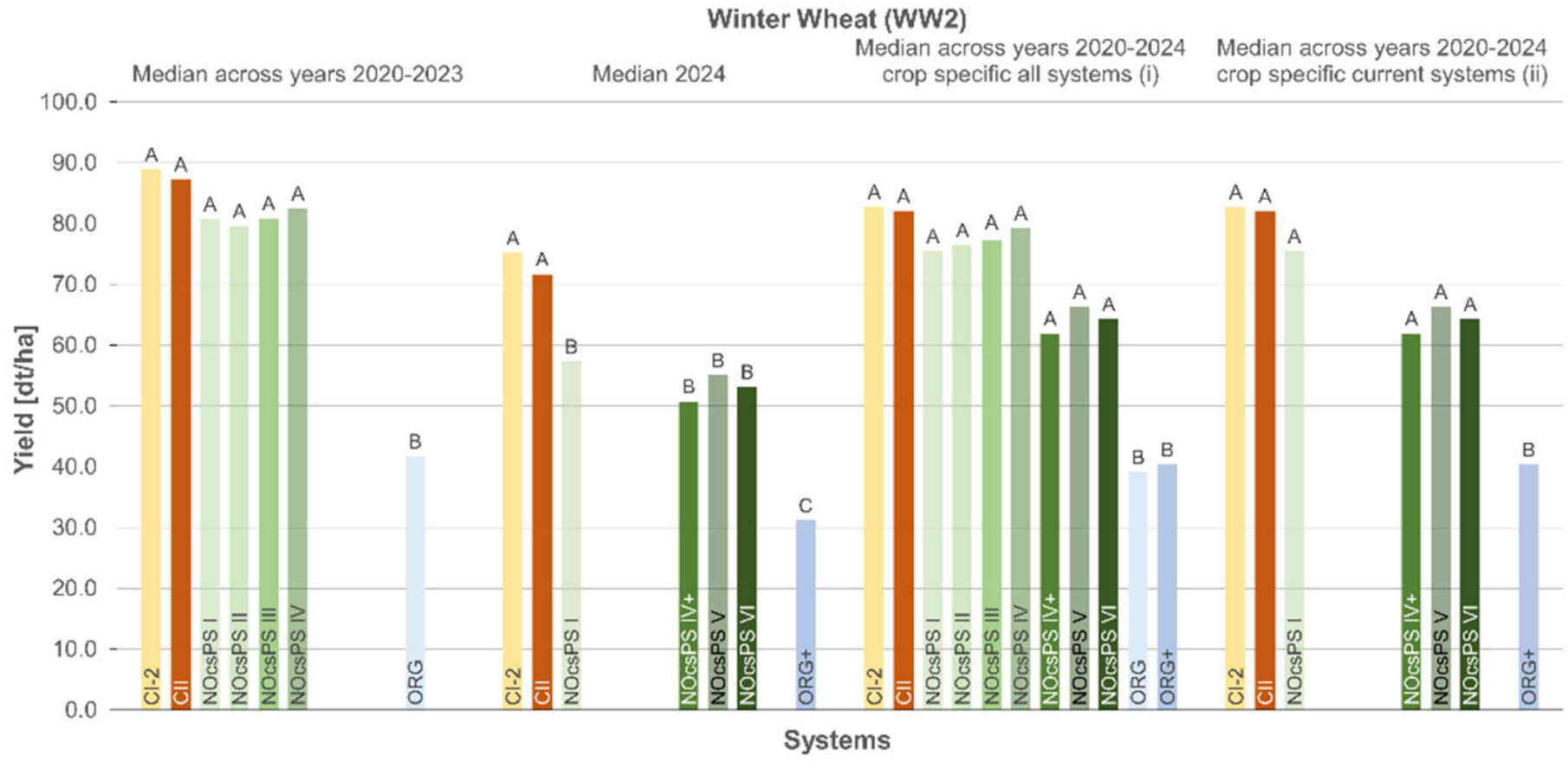

**Figure 3**: Medians of winter wheat yields [dt/ha] across different timeframes, years 2020-2023, year 2024 and years 2020-2024 of the tested systems for RGT Reform in Hohenheim. Letter display for mean comparison among systems for timeframe 2020-2023 data based on Model 4 and for timeframe 2020-2024 data based on Model 5 for (i) all systems and for (ii) the current Systems CI-2, CII, NOcsPS I, NOcsPS V, NOcsPS VI, NOcsPS IV+ and ORG+ only.

**Table 10**: Arithmetic means of winter wheat yields [dt/ha] (RGT Reform, Hohenheim) of the single years 2020 to 2024, across years 2020-2023 and 2020-2024 for all tested systems.

| Year | **CI-2** | **CII** | **NOcsPS I** | NOcsPS II | NOcsPS III | NOcsPS IV | **NOcsPS IV+** | **NOcsPS V** | **NOcsPS VI** | ORG | **ORG+** |
|---|---|---|---|---|---|---|---|---|---|---|---|
| 2020 | 93.3 | 93.9 | 81.9 | 78.7 | 83.0 | 86.0 | | | | 38.3 | |
| 2021 | 77.0 | 80.0 | 76.2 | 74.4 | 77.6 | 79.9 | | | | 47.2 | |
| 2022 | 95.3 | 90.7 | 93.1 | 93.5 | 92.9 | 97.1 | | | | 48.9 | |
| 2023 | 70.4 | 74.3 | 69.6 | 70.3 | 67.8 | 68.3 | | | | 32.8 | |
| Across years 2020-2023 | 88.5 | 88.2 | 80.2 | **79.3** | **80.3** | **81.2** | | | | 41.8 | |
| 2024 | 73.5 | 71.7 | 57.4 | | | | 50.9 | 55.1 | 53.2 | | 31.2 |
| Across years 2020-2024 | **84.8** | **84.1** | **75.7** | | | | | | | | |

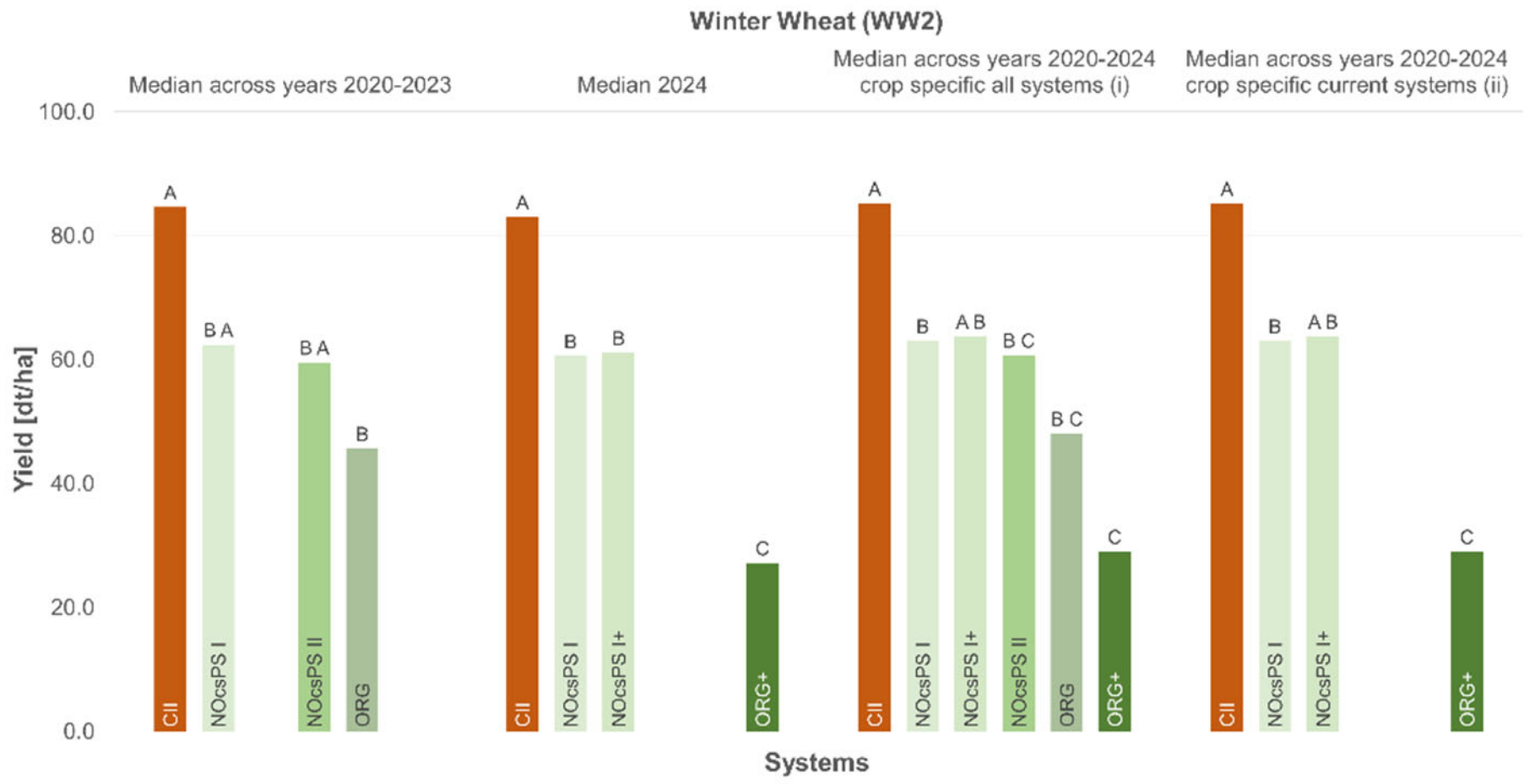

**Figure 4**: Medians of winter wheat yields [dt/ha] across different timeframes, years 2020-2023, year 2024 and years 2020-2024 of the tested systems, used varieties Govelino (System ORG in 2020) and RGT Reform (over all other systems and years) in Dahnsdorf. Letter display for mean comparison among systems for timeframe 2020-2023 data based on Model 4 and for timeframe 2020-2024 data based on Model 5 for (i) all systems and for (ii) the current Systems CII, NOcsPS I, NOcsPS I+ and ORG+ only.

**Table 11**: Arithmetic means of winter wheat yields [dt/ha] (RGT Reform, Dahnsdorf and variety Govelino for System ORG in 2020) of the single years 2020 to 2024, across years 2020-2023 and 2020-2024 for all tested systems.

| Year | CII | NOcsPS I | NOcsPS I+ | NOcsPS II | ORG | ORG+ |
|---|---|---|---|---|---|---|
| 2020 | 93.3 | 82.1 | | 78.9 | 44.3 | |
| 2021 | 81.3 | 62.8 | | 55.0 | 53.0 | |
| 2022 | 81.8 | 58.5 | | 60.3 | 48.2 | |
| 2023 | 87.5 | 55.0 | | 55.6 | 36.4 | |
| Across years 2020-2023 | 86.0 | 64.6 | | **62.4** | 45.5 | |
| 2024 | 83.1 | 60.8 | 61.2 | | | 27.5 |
| Across years 2020-2024 | **85.4** | **63.8** | | | | |

In contrast to the decreases in yield for winter wheat in 2024, in Hohenheim maize yields increased in 2024 in all tested systems compared to the mean average for the years 2020 to 2023, indicating beneficial conditions for maize in 2024 (Figure 5 and Table 12). It should also be noted that maize in the newly designed systems (NOcsPS IV+, NOcsPS V and NOcsPS VI) may have benefited more than other crops from the use of organic fertilizers. In contrast to Hohenheim, maize yields in Dahnsdorf declined in all cropping systems in 2024 compared to 2020-2023 (Figure 6 and Table 13).

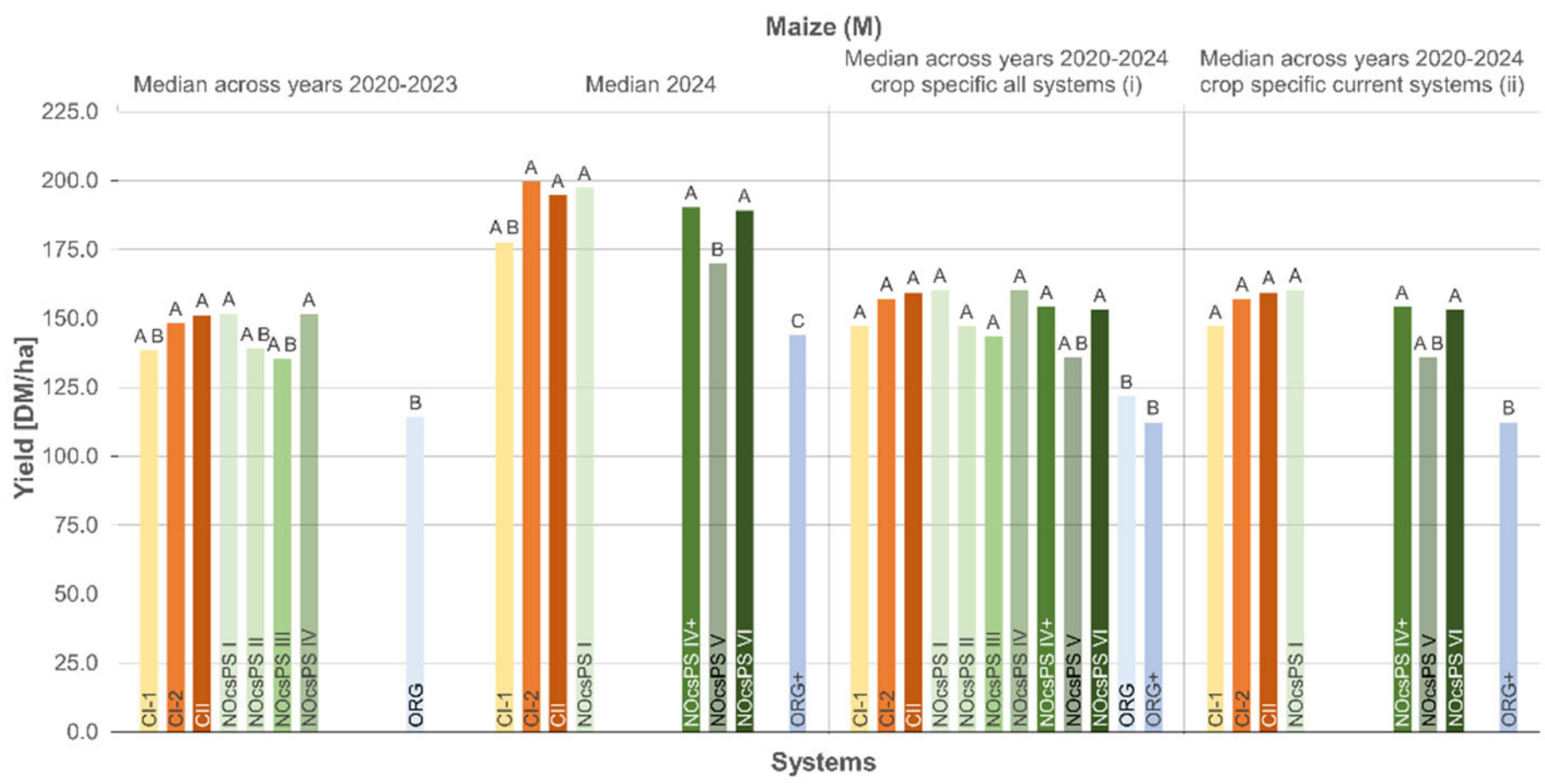

**Figure 5**: Medians of maize yields [DM/ha] across different timeframes, years 2020-2023, year 2024 and years 2020-2024 of the tested systems for Ronaldinio and LG 32257 in 2024 for Hohenheim. Letter display for mean comparison among systems for timeframe 2020-2023 data based on Model 4 and for timeframe 2020-2024 data based on Model 5 for (i) all systems and for (ii) the current Systems CI-I, CII, NOcsPS I, NOcsPS V, NOcsPS VI, NOcsPS IV+ and ORG+ only.

**Table 12**: Arithmetic means of maize yields [DM/ha] (Ronaldinio and LG 32257 in 2024, Hohenheim) of the single years 2020 to 2024, across years 2020-2023 and 2020-2024 for all tested systems.

| Year | CI-1 | CI-2 | CII | NOcsPS I | NOcsPS II | NOcsPS III | NOcsPS IV | NOcsPS IV+ | NOcsPS V | NOcsPS VI | ORG | ORG+ |
|---|---|---|---|---|---|---|---|---|---|---|---|---|
| 2020 | 105.2 | 122.7 | 119.3 | 124.6 | 90.1 | 95.5 | 108.5 | | | | 116.6 | |
| 2021 | 190.9 | 204.2 | 207.4 | 169.5 | 182.5 | 174.6 | 199.7 | | | | 105.2 | |
| 2022 | 137.7 | 148.0 | 153.3 | 159.9 | 150.4 | 141.2 | 161.8 | | | | 120.7 | |
| 2023 | 131.2 | 126.9 | 133.1 | 156.8 | 142.7 | 140.1 | 143.9 | | | | 115.6 | |
| Across years 2020-2023 | 141.2 | 150.5 | 153.3 | 152.7 | 141.5 | 137.8 | 153.5 | | | | 114.5 | |
| 2024 | 178.1 | 199.6 | 195.1 | 198.0 | | | | 191.1 | 170.6 | 189.7 | | 145.9 |
| Across years 2020-2024 | **148.6** | **160.3** | **161.6** | **161.8** | | | | | | | | |

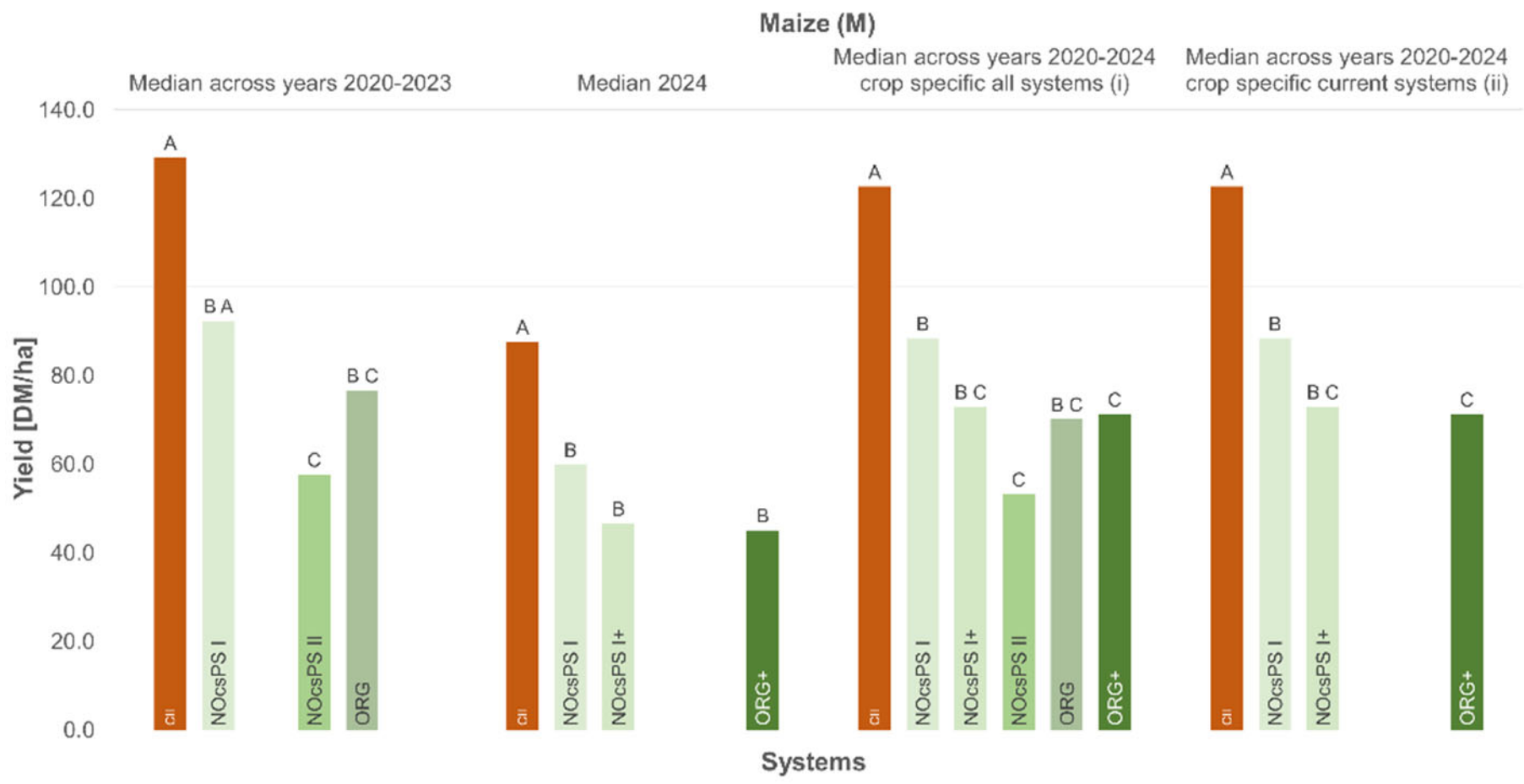

**Figure 6**: Medians of maize yields [DM/ha] across different timeframes, years 2020-2023, year 2024 and years 2020-2024 of the tested systems for Ronaldinio for Dahnsdorf. Letter display for mean comparison among systems for timeframe 2020-2023 data based on Model 4 and for timeframe 2020-2024 data based on Model 5 for (i) all systems and for (ii) the current Systems CII, NOcsPS I, NOcsPS I+ and ORG+ only.

**Table 13**: Arithmetic means of maize yields [DM/ha] (Ronaldinio, Dahnsdorf) of the single years 2020 to 2024, across years 2020-2023 and 2020-2024 for all tested systems.

| Year | CII | NOcsPS I | NOcsPS I+ | NOcsPS II | ORG | ORG+ |
|---|---|---|---|---|---|---|
| 2020 | 137.9 | 60.6 | | 69.0 | 81.9 | |
| 2021 | 126.0 | 119.6 | | 50.9 | 90.8 | |
| 2022 | 98.9 | 70.5 | | 39.9 | 72.6 | |
| 2023 | 180.7 | 152.8 | | 87.5 | 68.7 | |
| Across years 2020-2023 | 135.9 | 100.9 | | **61.8** | 78.5 | |
| 2024 | 88.6 | 65.3 | 51.9 | | | 46.2 |
| Across years 2020-2024 | **126.4** | **93.7** | | | | |

In 2024, legume yields at both sites exceeded the average yields for 2020–2023 in all systems (Figure 7 and Figure 8). With regard to soybeans in Hohenheim (Figure 7 and Table 14), the yields of all current NOcsPS systems do not differ significantly from the yields of conventional systems on a five-year average. The fact that the NOcsPS II and III systems show significantly lower yields on a five-year average can be seen as confirmation that the decision to discontinue these systems was the right one. The same conclusion can be drawn for peas in Dahnsdorf (Figure 8 and Table 15), as the modified systems NOcsPS I+ and ORG+ perform better on average over five years than the discontinued systems NOcsPS III and ORG.

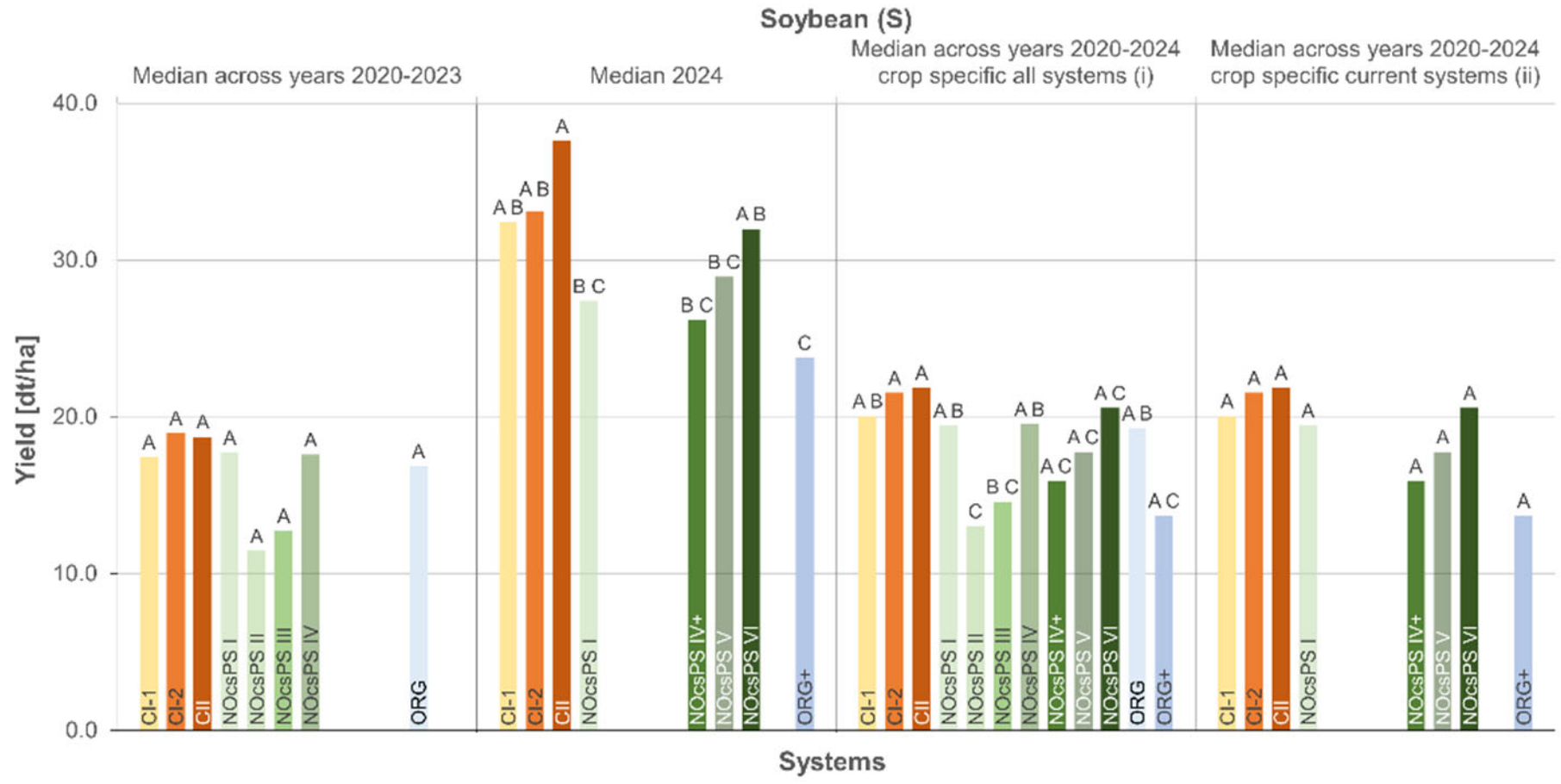

**Figure 7**: Medians of soybean yields [dt/ha] across different timeframes, years 2020-2023, year 2024 and years 2020-2024 of the tested systems for Sculptor in Hohenheim. Letter display for mean comparison among systems for timeframe 2020-2023 data based on Model 4 and for timeframe 2020-2024 data based on Model 5 for (i) all systems and for (ii) the current Systems CI-1, CII, NOcsPS I, NOcsPS V, NOcsPS VI, NOcsPS IV+ and ORG+ only.

**Table 14**: Arithmetic means of soybean yields [dt/ha] (Sculptor, Hohenheim) of the single years 2020 to 2024, across years 2020-2023 and 2020-2024 for all tested systems.

| Year | **CI-1** | **CI-2** | **CII** | **NOcsPS I** | NOcsPS II | NOcsPS III | NOcsPS IV | **NOcsPS IV+** | **NOcsPS V** | **NOcsPS VI** | ORG | **ORG+** |
|---|---|---|---|---|---|---|---|---|---|---|---|---|
| 2020 | 4.1 | 6.6 | 4.8 | 10.5 | 0.5 | 0.9 | 11.4 | | | | 14.1 | |
| 2021 | 34.7 | 34.8 | 29.7 | 23.2 | 24.8 | 26.3 | 22.9 | | | | 25.2 | |
| 2022 | 14.4 | 14.9 | 17.5 | 18.1 | 15.5 | 15.1 | 16.3 | | | | 16.5 | |
| 2023 | 27.5 | 28.0 | 30.1 | 20.9 | 18.7 | 15.6 | 20.8 | | | | 13.6 | |
| Across years 2020-2023 | 20.5 | 21.1 | 20.9 | 18.2 | 14.7 | 15.5 | 18.8 | | | | 17.3 | |
| 2024 | 32.1 | 34.0 | 37.6 | 27.6 | | | | 27.3 | 29.6 | 32.9 | | 23.7 |
| Across years 2020-2024 | **20.1** | **21.5** | **21.9** | **19.4** | | | | | | | | |

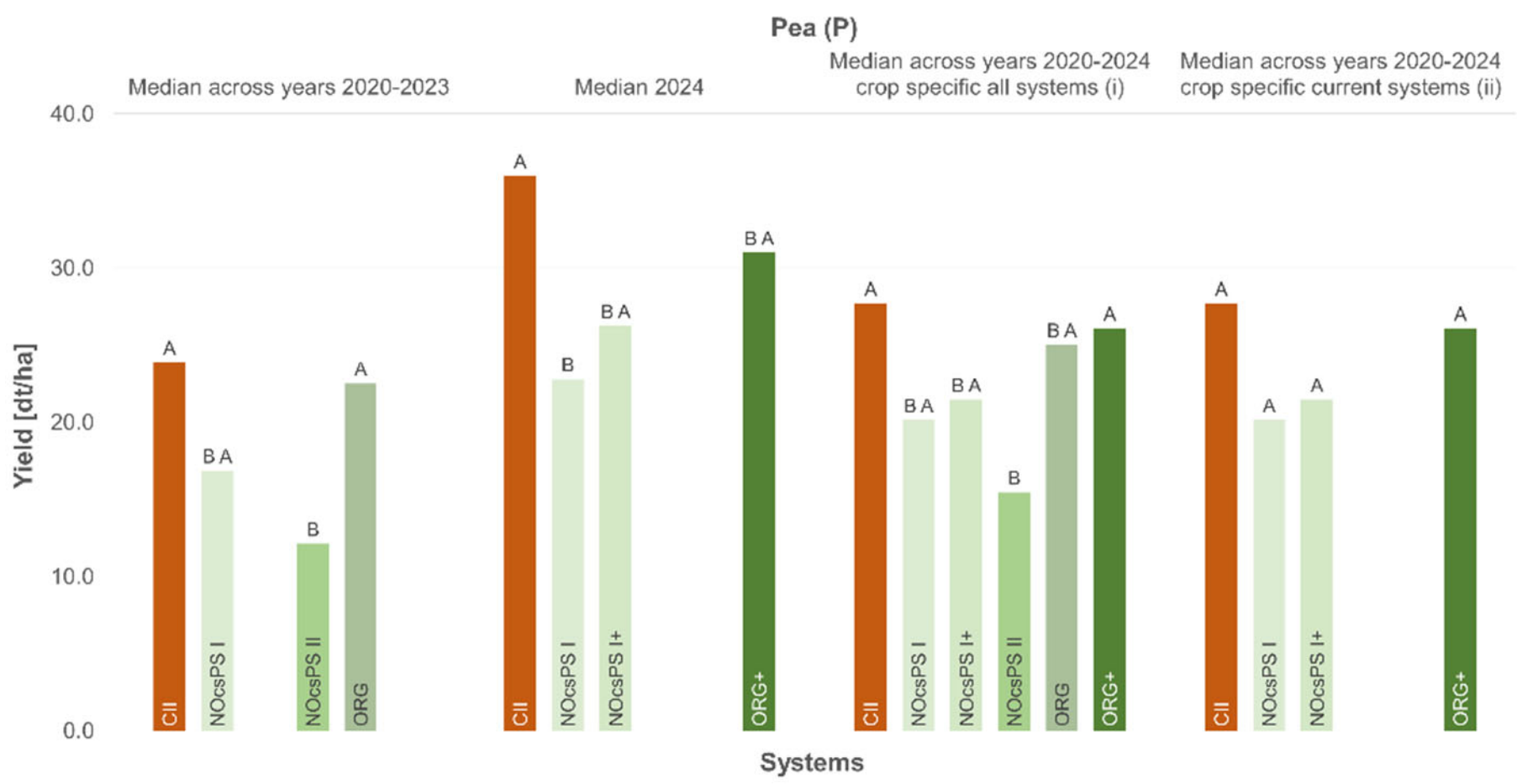

**Figure 8**: Medians of pea yields [dt/ha] across different timeframes, years 2020-2023, year 2024 and years 2020-2024 of the tested systems for Astronaut for Dahnsdorf. Letter display for mean comparison among systems for timeframe 2020-2023 data based on Model 4 and for timeframe 2020-2024 data based on Model 5 for (i) all systems and for (ii) the current Systems CII, NOcsPS I, NOcsPS I+ and ORG+ only.

**Table 15**: Arithmetic means of pea yields [dt/ha] (Astronaut, Dahnsdorf) of the single years 2020 to 2024, across years 2020-2023 and 2020-2024 for all tested systems.

| Year | **CII** | **NOcsPS I** | **NOcsPS I+** | NOcsPS II | ORG | **ORG+** |
|---|---|---|---|---|---|---|
| 2020 | 43.1 | 44.2 | | 27.3 | 46.4 | |
| 2021 | 26.8 | 21.0 | | 18.2 | 22.8 | |
| 2022 | 23.4 | 15.9 | | 16.8 | 19.4 | |
| 2023 | 16.9 | 7.5 | | 3.9 | 13.2 | |
| Across years 2020-2023 | 27.6 | 22.2 | | **16.6** | 25.4 | |
| 2024 | 36.0 | 23.0 | 26.3 | | | 31.1 |
| Across years 2020-2024 | **29.3** | **22.3** | | | | |

## Discussion

This paper proposes an approach based on linear mixed models to analyze data from a field trial with cropping systems subject to some changes of some of the systems after some years of trial. A key feature of the approach we suggest is to model year effects as fixed so that only within-year information is used. This strategy, adopted from established procedures for network meta-analysis, may be adaptively modified as more data from more years accrues. Specifically, the use of the Kenward-Roger method allows checking in each year if a transition to random year effects, and hence a recovery of inter-year information is worthwhile.

The results presented for the years 2020-2024 must be regarded as preliminary because only one year of data is available for the updated set of cropping systems. Going forward, it needs to be explored whether a washout period is required for the modified systems to account for carry-over effects (RECKLING et al., 2021). This would require either discarding data obtained during the washout period, or extending the model with carry-over effects. The decision as to the best modelling approach requires several more years and continuous monitoring of observed carry-over effects in each new year.

The comparison of the presented results for the Hohenheim and Dahnsdorf trials clearly shows how difficult it is to change the cropping system in a long-term trial. The modifications to the systems, such as the decision not to fertilize with the Cultan technique, the decision not to sow almost uniformly, the changed row spacing or the use of organic fertilizer, are associated with the expectation of achieving an improvement in productivity in the modified (current) systems. The effects of these changes can be investigated by comparing all systems and across all years of the trial. Presently, the available data does not yet allow any conclusions concerning the system changes made. It can be assumed that changes in yields in the different systems will become apparent only after further experimental years.

However, the methodological approach of the model, which allows the results of the systems of the first project phase with their realignment (NOcsPS IV+, NOcsPS V,NOcsPS VI and ORG+ in Hohenheim and NOcsPS I+ and ORG+ in Dahnsdorf) to be integrated into the second project phase, proves to be a valid solution for comparing the current systems or all systems tested over the years, whereby the significance of the evaluation opportunities of the described model can become apparent.

## Conclusion

Long-term experiments comparing different cropping systems may occasionally require a modification of some of the systems. Such changes require careful attention when it comes to joint analysis over years comprising several periods with different subsets of the systems being tested side-by-side. Using network meta-analytic concepts, linear mixed models can be formulated for this purpose, providing the opportunity to compare systems indirectly even if they were tested in different periods.

## Acknowledgements

This research was funded by Bundesministerium für Bildung und Forschung (BMBF), grant number 031B0731A. Hans-Peter Piepho was additionally supported by DFG project PI 377/24-1.

## Appendix: Calculation of adjusted means based on Model 3

Fitting Model 3 to the toy data in Table 3 by the method of least squares yields the effect estimates given in Table A1. With these effect estimates, the observations in the cells of a systems-by-years table can be predicted for all system-by-year combinations by adding the estimates of the intercept and the effects of the corresponding system and year given in Table A1. Importantly, this can be done even for those cells for which there is no observed data. Table A2 shows these predictions for the toy data. Taking simple averages over the cell means per system yields the adjusted means for the systems based in Model 3 (also see Table 4).

**Table A1**: Least squares estimates of effects in Model 3 for toy data in Table 3.

| Effect | Estimate |
|---|---|
| Intercept | 63.000 |
| System: ORG | –4.125 |
| System: C II | 35.100 |
| System: NOcsPS I | –0.100 |
| System: NOcsPS II | –0.875 |
| System: ORG+ | 4.000 |
| System: NOcsPSI+ | 0 |
| Year: 2020 | −6.750 |
| Year: 2021 | −9.500 |
| Year: 2022 | −8.000 |
| Year: 2023 | −9.250 |
| Year: 2024 | 0 |

**Table A2**: Predicted cell means (Intercept + System + Year) for toy data based on least-squares fit in Table A1.

| System | Year | | | | | System mean |
|---|---|---|---|---|---|---|
| | 2020 | 2021 | 2022 | 2023 | 2024 | |
| ORG | 63.000 –4.125–6.750=52.125 | 63.000 –4.125–9.500=49.375 | 63.000 –4.125–8.000=50.875 | 63.000 –4.125–9.250=49.625 | 63.000 –4.125+0=58.875 | 52.175 |
| C II | 63.000+35.100–6.750=91.350 | 63.000+35.100–9.500=88.600 | 63.000+35.100–8.000=90.100 | 63.000+35.100–9.250=88.850 | 63.000+35.100+0=98.100 | 91.400 |
| NOcsPS I | 63.000 –0.100–6.750=56.150 | 63.000 –0.100–9.500=53.400 | 63.000 –0.100–8.000=54.900 | 63.000 –0.100–9.250=53.650 | 63.000 –0.100+0=62.900 | 56.200 |
| NOcsPS II | 63.000 –0.875–6.750=55.375 | 63.000 –0.875–9.500=52.625 | 63.000 –0.875–8.000=54.125 | 63.000 –0.875–9.250=52.875 | 63.000 –0.875+0=62.125 | 55.425 |
| ORG+ | 63.000 +4.000–6.750=60.250 | 63.000 +4.000–9.500=57.500 | 63.000 +4.000–8.000=59.000 | 63.000 +4.000–9.250=57.750 | 63.000 +4.000+0=67.000 | 60.300 |
| NOcsPS I+ | 63.000 +0 –6.750=56.250 | 63.000 +0 –9.500=53.500 | 63.000 +0 –8.000=55.000 | 63.000 +0 –9.250=53.750 | 63.000 +0 +0=63.000 | 56.300 |